# Hybrid continuum-molecular modeling of fluid slip flow

**Mohamed Shaat**

*Mechanical Engineering Department, Abu Dhabi University, Al Ain, P.O.BOX 1790, UAE*

Email: shaatscience@yahoo.com & mohamed.i@adu.ac.ae

**Abstract**

Experiments on fluid systems in micro-/nano-scale solid conveyors have shown a violation of the no-slip assumption that have been adopted by the classical fluid mechanics. To correct this mechanics for the fluid slip, various approaches have been proposed to determine the slip boundary conditions. However, these approaches have revealed contradictory results for a variety of systems, and a debate on the mechanisms and the conditions of the fluid slip/no-slip past solid surfaces is sustained for a long time. In this paper, we establish the hybrid continuum-molecular modeling (HCMM) as a general approach of modeling the fluid slip flow under the influence of excess fluid-solid molecular interactions. This modeling approach postulates that fluids flow over solid surfaces with/without slip depending on the difference between the applied impulse on the fluid and a drag due to the excess fluid-solid molecular interactions. In the HCMM, the Navier-Stokes equations are corrected for the excess fluid-solid interactions. Measures of the fluid-solid interactions are incorporated into the fluid's viscosity. We demonstrate that the correction of the fluid mechanics by the slip boundary conditions is not an accurate approach, as the fluid-solid interactions would impact the fluid internally. To show the effectiveness of the proposed HCMM, it is implemented for water flow in nanotubes. The HCMM is validated by an extensive comparison with over 90 cases of experiments and molecular dynamics simulations of different fluid systems. We foresee that the hybrid continuum-molecular modeling of the fluid slip flow will find many important implementations in fluid mechanics.

**Keywords:** hybrid continuum-molecular modeling; slip flow; paradoxes; nanofluidics; fluid-solid interactions.

## 1. Introduction

The majority of problems of fluid mechanics are concerned with solving the Navier-Stokes equations for the fluid system. The validity of these equations to solve a long list of simple and complex problems of fluid mechanics have been experimentally proved. However, problems that can be solved by Navier-Stokes equations postulate no-slip and no-interaction of the fluid system with the solid surface. Recently, experiments on fluid systems in micro-/nano-scale solid conveyors have shown violation of the no-slip assumption, and it has been apparently shown that the Navier-Stokes equations incorrectly represent the mechanics of such fluid systems[1–13]. For example, the fluid flow rates in nanostructures, e.g., nanotubes and nanochannels, were observed with extremely higher values than usual[2,6,8]. This has been attributed to the molecular interactions between the fluid molecules and the molecules of the solid surface[14–18]. These excess interactions motivate the fluid slip over the solid surface and anomalous fluid flow.

With the aim to correct the Navier-Stokes equations for the fluid slip, various approaches have been proposed to determine the slip boundary conditions. Experimental and molecular dynamics approaches have been used to determine the slip length, which is the distance behind the fluid-solid interface at which the fluid velocity extrapolates to zero[19]. This has been calculated for the experimental measurements of the dynamic drainage force at different distances from the solid surface[19–21] and the experimental measurements of the flow profile near the fluid-solid interface[2,8,19]. However, these efforts have revealed contradictory results for a variety of systems, and a debate on the exact value of the slip length is sustained for a long time[14,19,22,23].

The debate on describing the nature of boundary conditions at the fluid-solid interface can be dated back to the 19th century where the pioneers of the current form of the fluid mechanics have expressed different opinions on this matter[24,25]. Thus, various hypotheses were proposed to describe the fluid slip over solid surfaces[26]. One hypothesis suggested that the fluid layer that is in contact with the solid surface exhibits no motion relative to it, and one portion of the fluid slips past the fluid portion adjacent to the solid surface[27]. Another hypothesis came to suggest that the few fluid layers adjacent to the solid surface are strongly held to it by molecular interactions[28]. According to the latter hypothesis, the slip would take place either directly at the solid surface or internal between one fluid portion past the other.

With the development of the surface force apparatus and atomic force microscopy, the experimental nanoscale probing of the fluid characteristics became possible, and it has been used to demonstrate the first hypothesis that suggests no-slip boundary conditions for fluids near hydrophilic surfaces, e.g., mica[29,30]. However, recent experimental studies that used other techniques have reported cases of fluid slip past solid surfaces[2,6–8]. These new studies indicated that the second hypothesis is more applicable, and the slippage of the fluid is more likely to be dependent on the molecular interactions of the fluid particles with the solid particles. However, many mechanisms that impact the fluid flow are still unresolved by these two simple hypotheses.

Here, we propose a hypothesis that fully describes the mechanisms by which the fluid interacts with solid surfaces and exhibits slip/no-slip past these surfaces. Based on this hypothesis, we evaluate the various models of the slip boundary conditions. We demonstrate that the correction of the classical fluid models for the fluid slip by the slip boundary conditions is not an accurate approach. This approach breaks down for fluid systems that involve high fluid-solid molecular interactions, e.g., fluids in nanostructures.

In addition, we establish the hybrid continuum-molecular modeling as a general approach of modeling the fluid slip flow under the influence of excess fluid-solid molecular interactions. In the context of this approach, the Navier-Stokes equations are corrected for the excess fluid-solid interactions by incorporating measures of these interactions into the fluid's viscosity. The hybrid continuum-molecular modeling is an effective approach, as it permits the implementation of the conventional no-slip boundary conditions. To show the applicability of this approach, a hybrid-continuum molecular model of water flow in nanotubes is developed. This model is validated by a comparison with over 90 cases of experiments and molecular dynamics simulations of different fluid systems.

## 2. Fluid-Solid Interface: Mechanisms of Slip/No-Slip

In this section, a new hypothesis that fully describes the mechanisms by which the fluid approaches and flows over solid surfaces with/without slip is proposed. This hypothesis depends on breaking the fluid into a set of layers (Fig. 1). Each layer is linked to the solid surface by a molecular attraction force. The strength of this force depends on the location of the fluid layer from the solid surface. It is maximum at the first fluid layer and decreases for the other fluid layers. When the fluid is subjected to a driving force, all layers are subjected to the same impulse. If the impulse given to the fluid layer by the driving force is higher than the dragging-impulse (or drag) associated with the fluid-solid molecular attraction of this layer, the fluid layer apparently moves with respect to the solid surface with a momentum that depends on the difference between the two impulses. However, if the applied impulse is lower than the drag associated with the fluid-solid attraction, the fluid layer exhibits no-motion with respect to the solid surface. The fluid slips over the solid surface if the momentum given to the first fluid layer (or the impulse acting on the first fluid layer) is higher than the drag due to the fluid-solid attraction, and the slip velocity depends on the difference between the applied impulse and the drag.

The fluid approaches the solid surface *with no direct contact*. Instead, a very thin region that separates the first fluid layer and the solid surface may exist. This region is commonly known as the depletion region[16,18,31–35]. The thickness of this region depends on the wettability of the surface[18,34]. If the solid surface is hydrophobic, the fluid-surface interactions are repulsive forces at the surface, and the fluid particles are pushed away from the solid surface forming the depletion region. The pushed fluid particles accumulate the first fluid layer (may accumulate the other subsequent layers too) after the depletion region (Fig. 1). This increases the fluid density at this layer. Under the same momentum, the increase in the density of the fluid layer decreases its velocity. For hydrophilic solid surfaces, the depletion layer may vanish, and the fluid particles would be in a direct contact with the solid surface. The direct contact to the solid surface does not guarantee no-slip of the first fluid layer past the solid surface. This mainly depends on the applied impulse relative to the drag due to the fluid-solid attraction.

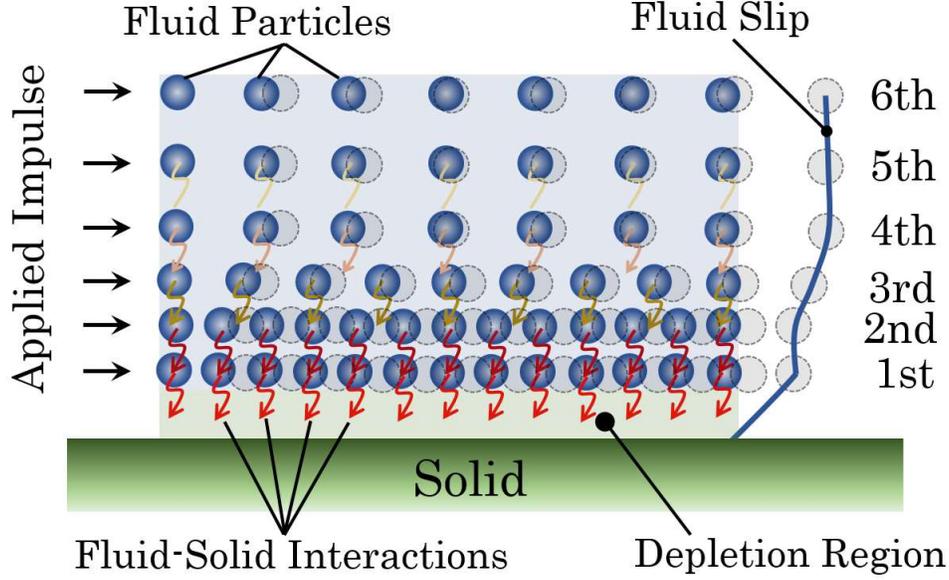

**Figure 1: Hypothesis of Fluid Slip over Solid Surfaces.** The fluid is represented as a set of layers, which are subjected to the same applied impulse. Each fluid layer is linked to the solid surface by fluid-solid attraction forces, which produce a drag to the fluid layer flow. The fluid layers slip with respect to the solid surface as well as to each other. The slip of a fluid layer depends on the difference between the applied impulse and the drag.

Based on the proposed hypothesis, the first fluid layer (and probably the few subsequent layers) would exhibit no-slip if the applied impulse does not be able to move this dense layer and/or does not be able to outweigh the drag due to the attraction between this fluid layer and the solid surface. At the subsequent layers, however, the strength of the attraction force and the density decrease, and the applied impulse would be big enough to push these fluid layers to move with respect to the solid surface. The fluid layers far from the solid surface are of lower densities and weakly held to the solid surface, and, therefore, these layers would exhibit a plug flow (Fig. 1). This indicates that the few fluid layers near the solid surface are generally dragged with either slip or no-slip with respect to the solid surface.

## 3. Paradoxes of the Slip Boundary Conditions

The recent observations of the fluid slip past solid surfaces motivated many scholars who reported the slip length that can characteristically describe the fluid slip at the solid boundary. The reported values of the slip length, $L_s$, depended on Navier's linear boundary condition[26,36]:

$$v_s = L_s \mathbf{n} \cdot (1 - \mathbf{nn}) \cdot (\nabla \mathbf{v} + (\nabla \mathbf{v})^T) \tag{1}$$

where $\mathbf{v}$ is the velocity vector, and $v_s$ is the component of the fluid velocity tangent to the solid surface (the slip velocity). $\mathbf{n}$ is the unit normal vector to the solid surface.

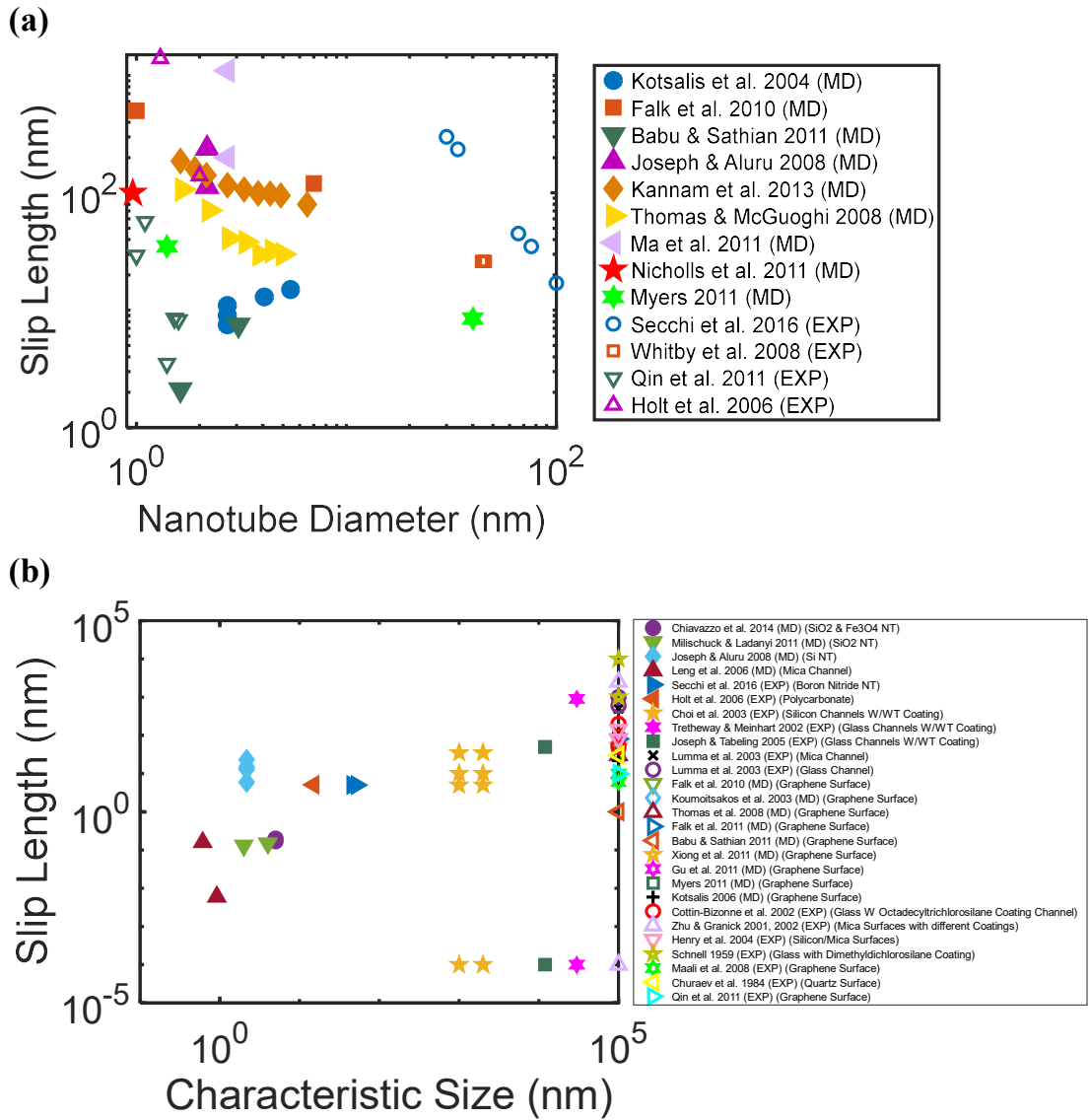

**Figure 2:** A review of the slip length values determined by experiments (EXP) and molecular dynamics (MD) of various water systems. (a) The slip length of water in carbon nanotubes (CNTs) versus the nanotube diameter. (b) The slip length of water on various flat surfaces and water in various nanotubes (NT) versus the characteristic size of water confinement.

The Navier's linear boundary condition (Eq. (1)) has been implemented in various experimental methods (reviewed in[26]) and molecular dynamics simulations that reported the slip length of different fluid systems. However, this boundary condition does not hold in all fluid systems. This model ignores one essential measure, which is the cause of the fluid slip. According to the proposed hypothesis, the fluid slip mainly depends on the fluid-solid interactions. Nonetheless, these interactions are totally disregarded by the slip length model in Eq. (1). The issue that resulted in a growing debate on the exact value of the slip length[14,19,22,23]. High discrepancies between the reported values of the slip length in the literature can be observed. With the aim to reveal the reasons that have led to these discrepancies, we reviewed in Fig. 2 the values of the slip length of various fluid systems reported in the literature.

The experimental debate on the exact value of the slip length is evidenced by noticing the wide spectrum of the slip length values reported by the different experimental methods (Fig. 2). The early experimental methods of probing the fluid properties and its flow characteristics near flat surfaces proved the no-slip of water on mica and glass hydrophilic flat surfaces[29,30,37]. Nonetheless, high slip lengths of 1000 nm[5] and 2500 nm[3,13,21] were also experimentally reported when the wettability of these surfaces were altered by coatings. In addition, the slip length of water on graphene[6,38] and quartz[39] flat surfaces was reported by 10 nm and 30 nm, respectively. Despite the changes in the surface wettability are not expected to reflect significantly contradictory values of the slip length, these experiments exhibited paradoxical results.

Because probing the fluid characteristics in micro/nano-channels and nanotubes is challenging, experiments on fluids in these nanostructures depended on measuring the fluid flow characteristics by collecting the fluid's permeate[2,8] or tracing the fluid's flow by means of field effect transistors[6]. These experimental techniques also gave discrepancies in the reported slip conditions. For instance, the slip length of water in silicon microchannels of $1 - 2$ μm depth was reported by $0 - 10$ nm[40]. The slip was slightly increased to $5 - 35$ nm upon coating the silicon surface of the same channels by octadecyltrichlorosilane. In one study, a glass microchannel for water revealed no-slip[10] while another study[11] reported slip lengths of 1000 nm and 860 nm for water in glass and mica channels, respectively. The slip length for water in octadecyltrichlorosilane coated-glass microchannels was reported by 0 and 900 nm in two different studies[9,10]. As for water in carbon nanotubes (CNTs), discrepancies are clearly seen in Fig. 1. Majumder et al.[8] reported a huge value of the slip length of 39 mm – 68 mm for water in 7 nm diameter-CNT. Then, Holt et al.[7] reported 140 nm and 1400 nm slip length values for 2 nm and 1.3 nm CNTs, respectively. In addition, slip length values of 26 nm, $8 - 53$ nm, and $17 - 300$ nm were reported for water in CNTs with diameters of $44 - 46$ nm, $0.81 - 1.59$ nm, and $30 - 100$ nm, respectively[1,2,6]. In contrary, the experiment by Sinha et al.[41] gave no-slip of water in 200-300 nm CNTs.

The slip conditions as reported by the molecular dynamics (MD) simulations are also discrepant. The slip length of water on a graphene sheet was reported between 1 nm and 112 nm in various studies[32,42–47]. In addition, the slip length of water in CNTs with $1 - 2.71$ nm diameters was reported between 2 nm[42] and 500 nm[44]–1000 nm[48].

**3.1. Reasons of paradoxes**

Based on the proposed hypothesis of the fluid flow under the influence of fluid-solid interactions, the various models of the fluid slip are evaluated (these models are reviewed in Table 1). Staring with the slip length as defined in Eq. (1), this model assumes a linear extrapolation of the velocity profile of a fluid with constant viscosity (Fig. 3(a)). The assumed velocity profile by this model has a velocity jump at the fluid-solid interface followed by the velocity profile that is usually obtained by the no-slip fluid model. Thus, for a fluid flow in a circular tube (for example), the velocity profile is parabolic in the fluid core with a velocity jump at the tube wall (Fig. 3(a)). According to Eq. (1), the slip length depends on the slope of the velocity at the solid surface, which would vary from zero (for plug-like flow) to $pr/2\mu$ where $\mu$ is the fluid's viscosity, $p$ is the pressure gradient, and $r$ is the radial position. Consequently, the slip length would vary from zero for the no-slip flow (where $v_s = 0$) to infinity for the plug-like flow. This explains the wide range of the slip length reported in the literature. However, this simple model of the slip length (or even its modified version that depends on the nonlinear extrapolation[49]) fails to reveal the exact mechanisms of the fluid flow and the slip conditions. Because of the interactions between the fluid particles and the particles of the solid surface, the actual fluid flow is generally anomalous (Fig. 3(b)). For example, the fluid flow in tubes is generally non-parabolic, and the velocity profile has a jump at the interface that is followed by a drag at the first fluid layer[17] (Fig. 3(b)) (this can be understood based on the explained hypothesis).

The slip length in Eq. (1) depends on the slope of the velocity profile at the first fluid layer, $(\nabla \mathbf{v} + (\nabla \mathbf{v})^T)$. This slope is representative of the slope everywhere within the fluid domain. This is acceptable if the fluid is subjected to a constant shear stress where the velocity profile is described by a constant slope. However, for cases in which the fluid's shear stress is non-constant, e.g., fluid flow in tubes, the slope of the velocity profile is generally non-constant, as well. Therefore, a modification over the slip length model in Eq. (1) was proposed for fluid flow in tubes by considering the parabolic-nonlinear extrapolation of the velocity profile to calculate the slip length[48,49].

Another drawback of the slip length model in Eq. (1) is that it assumes the fluid with a constant viscosity. However, because the strength of the fluid-solid interactions changes from one fluid layer to the other, the viscosity is generally non-constant and varies depending on the separation from the solid surface[16,17]. Some models of the slip length that depend on an effective viscosity of the fluid were proposed[14,32,45]. The effective viscosity was calculated by the average of the interfacial viscosity to the bulk fluid's viscosity. Nonetheless, these models are not accurate enough, as they still assume constant viscosity of the fluid. These models disregard the spatial variation of the viscosity, which is mainly because of the spatial variation of the fluid-solid interactions. In addition, despite some other models (Table 1) represented the slip length dependent on the confinement size and/or the wettability of the solid surface, these models disregarded the spatial variation of the excess fluid-solid interactions and the viscosity. These models assumed a fluid flow, which – in many cases – is not the actual fluid flow under the influence of these excess interactions. This what have led to the discussed paradoxes.

Now, can we develop a slip length model that accounts for the spatial variation of the fluid-solid interactions? In fact, this is a challenging task, and – in many cases – it is impossible to derive such a model. The slip length being a boundary condition, it depends on the slope of the velocity profile at the boundary, and this localized value of the slope is a representative of the slope everywhere. However, this is not true, as the slope of the fluid velocity spatially varies due to the fluid-solid interactions. Thus, the spatial variation of the velocity slope cannot be captured by a boundary conditions, e.g., the slip length. The spatial variation of the velocity slope would require another measure, which reveals the influence of the fluid-solid interactions on the fluid's characteristics. As shown in Fig. 3, the spatial variation of the velocity slope can be effectively modeled by a spatially varying viscosity function. This can be understood by the Newton's law of viscosity that defines the viscosity as the ratio of the applied shear stress to the shear rate (or the slope of the velocity). When there is no fluid-solid interaction (Fig. 3(a)), the shear rate is proportionally related to the applied shear stress, and the constant of proportionality is the fluid's viscosity. However, the proportionality relation between the shear rate and the applied shear stress breaks down due to the fluid-solid interactions (Fig. 3(b)).

Next, we propose an effective model of the fluid slip flow under the influence of the fluid-solid interactions that can replace the existing models of the slip boundary conditions.

**Table 1:** Models of the slip length of fluid in tubes as reported in the literature.

| Slip Length Model | Assumptions | Ref. |
|---|---|---|
| Eq. (1) | - Linear extrapolation.<br>- Constant viscosity of water in nanotube.<br>- The velocity profile has a velocity jump at the interface followed by a parabolic flow (Fig. 2(a)). | 2 |
| $L_s(R) = 30 \text{ nm} + \dfrac{C}{(2R)^3}$<br>$C$ is a fitting parameter. | - Linear extrapolation.<br>- The velocity profile has a velocity jump at the interface followed by a parabolic flow (Fig. 2(a)).<br>- The slip length is a function of the nanotube radius, $R$.<br>- Constant water viscosity within the nanotube. However, an effective water viscosity that depends on the nanotube radius was considered.<br>- The effective viscosity is the weighted average of an arbitrary assumed interfacial viscosity of ~0.655 mPa·s and the bulk water viscosity. | 45 |
| $L_s = \left(\dfrac{\mu_0}{k}\right)\left(\dfrac{v_s}{v_0 \text{ arcsinh}\left(\dfrac{v_s}{v_0}\right)}\right)$<br>$k$ is the dynamic coefficient of friction. $v_0$ (m/s) is a fitting parameter. | - Linear extrapolation.<br>- Constant viscosity of water in nanotube.<br>- The velocity profile has a velocity jump at the interface followed by a parabolic flow (Fig. 2(a)).<br>- $L_s \geq 0$. | 48 |
| $L_s(R) = \delta\left(\dfrac{\mu_I}{\mu_c} - 1\right)\left[1 - \dfrac{3}{2}\dfrac{\delta}{R} + \left(\dfrac{\delta}{R}\right)^2 - \dfrac{1}{4}\left(\dfrac{\delta}{R}\right)^3\right]$ | - Linear extrapolation.<br>- Constant viscosity of water in nanotube.<br>- The velocity profile has a velocity jump at the interface followed by a parabolic flow (Fig. 2(a)).<br>- The slip length depends on the nanotube radius $R$, the thickness of the interface $\delta$, and the ratio of the interfacial viscosity ($\mu_I$) to the core viscosity ($\mu_c$). | 32 |
| $L_s(\theta) = \dfrac{C}{(\cos\theta + 1)^2}$<br>$C$ is a fitting parameter.<br>$\theta$ surface contact angle. | - Linear extrapolation.<br>- The velocity profile has a velocity jump at the interface followed by a parabolic flow (Fig. 2(a)).<br>- Constant water viscosity within the nanotube. However, an effective water viscosity that depends on the nanotube radius $R$ was considered.<br>- The effective viscosity is the weighted average of the interfacial viscosity to the bulk water viscosity.<br>- The interfacial viscosity ($\mu_I$) is related to the surface wettability:<br>$$\mu_I = \mu_0(-0.018\theta + 3.25)$$<br>- The slip length depends on the surface wettability. | 14 |

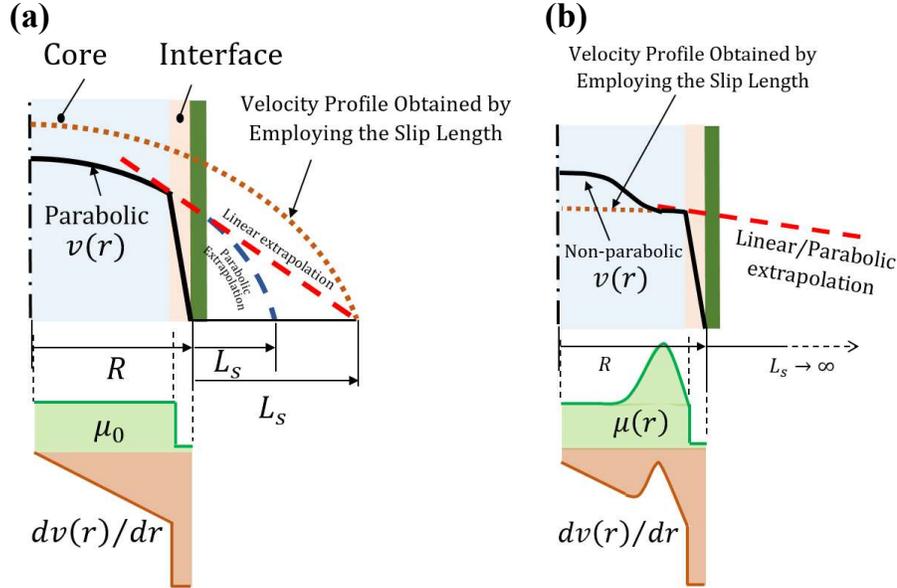

**Figure 3: Paradoxes of Slip Boundary Conditions.** (a) A model of fluid flow in tube that neglects the fluid-solid interactions. In this model, the fluid is assumed having a constant viscosity $\mu_0$ (green diagram), and, therefore, it flows with a parabolic velocity profile $v(r)$. At the interface, the fluid slips over the solid surface. The sip length is calculated to correct the velocity profile of this fluid model for the velocity jump at the interface. The slip length can be calculated by either the linear extrapolation (as in Eq. (1)) or by the parabolic extrapolation of the velocity profile. The slip length calculated by linear extrapolation overestimates the velocity profile of the fluid in tube. The linear extrapolation gives the slip length considering one slope of the velocity profile, $dv(r)/dr$ (brown), is representative of the fluid flow in the tube. The parabolic extrapolation, however, gives the slip length considering a spatial-linear variation of the slope of the velocity profile, $dv(r)/dr$ (brown). (b) A typical model of a fluid flow in tube that accounts for the fluid-solid interactions. Because of the fluid-solid interactions, the fluid flow is generally non-parabolic, the slope of the velocity profile spatially varies ($dv(r)/dr$, brown), and the viscosity has a radial distribution ($\mu(r)$, green). The slip length cannot extrapolate this case because neither the linear nor the parabolic extrapolations can give the same velocity profile. For an accurate representation of the fluid flow of this case, a spatially varying viscosity function, $\mu(r)$, should replace the slip boundary conditions.

## 4. Hybrid Continuum-Molecular Modeling of Fluid Slip Flow

The fluid flow near or between solid surfaces mainly depends on the way it interacts with the solid surface. The classical models of the fluid mechanics deal with the fluid in isolation of its confining solid surfaces. Nonetheless, the practical use of fluids necessitates the use of the fluid in a particular solid conveyor. Therefore, the properties and the flow characteristics of fluids should be explored considering measures of the fluid-solid interactions. Because the solid is located at and could only impact the fluid boundary, the fluid boundary conditions can be modified for the fluid-solid interactions (such as the fluid slip boundary conditions). Nonetheless, in many other cases, the fluid-solid interactions would exceed the boundary and impact the fluid internally, and the fluid properties are, therefore, altered by the fluid-solid interactions.

In the preceding sections, we demonstrated that the fluid-solid interactions significantly influence the fluid core as well as its boundary. In addition, we demonstrated that the slip boundary conditions can only reflect the influence of the fluid-solid interactions at the fluid's boundary; the issue that have led to the paradoxes of the slip boundary conditions. Here, we propose the hybrid continuum-molecular modeling (HCMM) as a general approach that can effectively reveal the influence of the fluid-solid interactions on the fluid characteristics. In the context of this approach, the fluid is a continuous liquid matter that undergoes various forms of polar or non-polar interactions with a solid surface. These interactions exceed the fluid boundary to significantly alter the fluid's properties. The HCMM gives more efficient and accurate modeling of the influence of the fluid-solid interactions on the fluid mechanics. The HCMM can replace the slip boundary conditions with new measures of the fluid properties. The interesting of the HCMM is that it allows the application of the no-slip boundary conditions, and, therefore, the paradoxes associated with the slip boundary conditions are avoided.

The HCMM is developed based on the proposed hypothesis of the fluid flow under the influence of fluid-solid attractions. A discrete system of fluid particles is considered. Each fluid particle is subjected to a set of local interactions with the other surrounding fluid particles. In addition, the fluid system undergoes interactions with a discrete system of solid particles forming the solid surface. Thus, each fluid particle is subjected to an excess molecular interaction force with the particles of the solid surface. The strength of this excess interaction depends on the distance between the fluid particle and the solid particle $|\mathbf{x} - \mathbf{x}_s|$, where $\mathbf{x}$ and $\mathbf{x}_s$ are the positions of the fluid and the solid particles, respectively. Accordingly, the Newtonian dynamic balance of a fluid particle $n$ can be expressed as follows:

$$\begin{aligned}\sum_{m=1}^{N_m} \mathbf{F}_{nm} + \boldsymbol{\mathcal{F}}_n &= \frac{\partial}{\partial t}(\widehat{m}\mathbf{v}) \\ \sum_{m=1}^{N_m} (\mathbf{x} \times \mathbf{F}_{nm}) + (\mathbf{x} \times \boldsymbol{\mathcal{F}}_n) &= \mathbf{x} \times \frac{\partial}{\partial t}(\widehat{m}\mathbf{v})\end{aligned} \quad (2)$$

where $\mathbf{F}_{nm}$ is the local interaction force between two fluid particles ($n$ and $m$). $\boldsymbol{\mathcal{F}}_n$ is the excess molecular-interaction force acting on the particle $n$ due to its interaction with the solid surface. $\mathbf{x} \times \mathbf{F}_{nm}$ and $\mathbf{x} \times \boldsymbol{\mathcal{F}}_n$ denote moment fields that generate due to the couple of the force fields $\mathbf{F}_{nm}$ and $\boldsymbol{\mathcal{F}}_n$ acting on the fluid particle. $N_m$ is the number of the fluid particles that surround the particle $n$. $\widehat{m}$ is the mass of the fluid particle, and $\mathbf{v}$ is the particles velocity vector.

The model of the discrete fluid particles can be extended to a fluid medium that occupies a volume $\Omega$ and is bounded by a surface $S$. According to Eq.(2), the continuity, balance of momentum, balance of moment of momentum equations can be, respectively, obtained as follows:

$$\frac{\partial}{\partial t}\left(\int_\Omega \widehat{m}\, d\Omega\right) = -\int_\Omega (\boldsymbol{\nabla} \cdot \widehat{m}\mathbf{v})\, d\Omega \quad (3)$$

$$\int_\Omega \mathbf{F}\, d\Omega + \int_\Omega \mathcal{F}\, d\Omega + \int_S \mathbf{T}\, dS = \int_\Omega \frac{\partial}{\partial t}(\widehat{m}\mathbf{v})\, d\Omega \qquad (4)$$

$$\int_\Omega (\mathbf{x}\times\mathbf{F})\, d\Omega + \int_\Omega (\mathbf{x}\times\mathcal{F})\, d\Omega + \int_S (\mathbf{x}\times\mathbf{T})\, dS = \int_\Omega \left(\mathbf{x}\times\frac{\partial}{\partial t}(\widehat{m}\mathbf{v})\right) d\Omega \qquad (5)$$

where **F** is a vector of the local body forces, which locally act on the fluid particle at point **x** due to its interactions with the surrounding fluid particles and any other local forces (such as gravitational and external forces) that would act on the fluid particle. $\mathcal{F}$ is a vector of the excess body forces that act on the fluid particle at point **x** due to the interaction of the fluid particle with the particles of the solid surface. The body force $\mathcal{F}$ represents any kind of polar and non-polar molecular interactions between the fluid particle and the solid particle. **T** is a vector of the fluid's surface tractions. The surface traction, **T**, generates various momentum fluxes acting on the fluid including the convective momentum fluxes and the molecular momentum fluxes. The viscous momentum flux usually represents the slip of one fluid layer past the other layers. In addition to this conventional momentum, a special viscous momentum that depends on the fluid-solid interactions is generated. Thus, a part of the applied surface traction **T** is consumed to give the fluid layer a momentum against its attraction to the solid surface (as explained in Section 2). To account for the excess momentum due to the fluid-solid interactions, the molecular momentum flux tensor is formed depending on two viscous momentum flux tensors, as follows:

$$\mathbf{T} = -(\mathbf{n}\cdot\boldsymbol{\phi})$$
with $\qquad\qquad\qquad\qquad\qquad\qquad\qquad\qquad\qquad\qquad\qquad\qquad\qquad\qquad(6)$
$$\boldsymbol{\phi} = P\mathbf{I} + \boldsymbol{\sigma} + \mathbf{t} + \rho\mathbf{v}\mathbf{v}$$

where **n** is the unit normal vector, and **I** is the unit dyadic. $\boldsymbol{\phi}$ is the combined momentum-flux tensor. $P$ is the applied pressure field. $\rho$ is the mass density of the fluid.

In Eq.(6), $\boldsymbol{\sigma}$ is the conventional viscous stress tensor, which is conjugate to the fluid-fluid interactions. **t** is a new viscous stress tensor that is conjugate to the fluid-solid interactions. It should be noted that the momentum-flux tensor and the viscous stress tensors presented in Eq.(6) are general tensors, which can be decomposed into symmetric and skew-symmetric parts. By substituting Eq. (6) into Eqs. (3)-(5) and applying the divergence theorem, the continuity and the balance equations of the fluid continuum can be written in the form:

$$\frac{\partial}{\partial t}\rho = -\boldsymbol{\nabla}\cdot\rho\mathbf{v} \qquad (7)$$

$$\frac{\partial}{\partial t}(\rho\mathbf{v}) = -\boldsymbol{\nabla}\cdot\rho\mathbf{v}\mathbf{v} - \boldsymbol{\nabla}P - \boldsymbol{\nabla}\cdot\boldsymbol{\sigma} - \boldsymbol{\nabla}\cdot\mathbf{t} + \rho\mathbf{g} \qquad (8)$$

$$\hat{\boldsymbol{\sigma}} + \hat{\mathbf{t}} = 0 \qquad (9)$$

where the gravitational force, $\rho\mathbf{g}$, is introduced as a body force. $\hat{\boldsymbol{\sigma}}$ and $\hat{\mathbf{t}}$ are stress vectors with the components $\hat{\sigma}_i = \epsilon_{ijk}\sigma_{jk}$ and $\hat{t}_i = \epsilon_{ijk}t_{jk}$, respectively.

Equation (9) indicates that the skew-symmetric parts of the viscous stress tensors, $\boldsymbol{\sigma}$ and $\mathbf{t}$, vanish. Thus, the balance equations (Eqs.(8) and (9)) can be rewritten as follows:

$$\frac{\partial}{\partial t}(\rho \mathbf{v}) = -\boldsymbol{\nabla} \cdot \rho \mathbf{v}\mathbf{v} - \boldsymbol{\nabla} P - \boldsymbol{\nabla} \cdot \boldsymbol{\sigma}^s - \boldsymbol{\nabla} \cdot \mathbf{t}^s + \rho \mathbf{g} \qquad (10)$$

where the balance equations depend on the symmetric part of the viscous stress tensors, $\boldsymbol{\sigma}^s = \frac{1}{2}(\boldsymbol{\sigma} + \boldsymbol{\sigma}^T)$ and $\mathbf{t}^s = \frac{1}{2}(\mathbf{t} + \mathbf{t}^T)$ where the superscript T stands for the transpose.

It should be mentioned that Eq.(10) represents a modified Navier-Stokes equation for the fluid-solid interactions. If the fluid-solid interactions are neglected (i.e., $\mathbf{t} = \mathbf{0}$), Eq.(10) reduces to the conventional Navier-Stokes equation.

The viscous stress tensors can be expressed based on Newton's law of viscosity as follows:

$$\boldsymbol{\sigma}^s = -\mu_0(\boldsymbol{\nabla}\mathbf{v} + (\boldsymbol{\nabla}\mathbf{v})^T) \qquad (11)$$

$$\mathbf{t}^s = -\mu_{sf}(\mathbf{x})(\boldsymbol{\nabla}\mathbf{v} + (\boldsymbol{\nabla}\mathbf{v})^T) \qquad (12)$$

where $\mu_0$ denotes the viscosity of the bulk fluid (the conventional viscosity), and $\mu_{sf}$ is a new viscosity that is introduced to model the fluid-solid interactions.

By combining the two stress tensors $\boldsymbol{\sigma}^s$ and $\mathbf{t}^s$ into an equivalent total viscous stress tensor $\boldsymbol{\tau} = \boldsymbol{\sigma}^s + \mathbf{t}^s$, such that:

$$\boldsymbol{\tau} = -\mu(\mathbf{x})(\boldsymbol{\nabla}\mathbf{v} + (\boldsymbol{\nabla}\mathbf{v})^T) \qquad (13)$$

Eq. (10) takes the conventional form of the Navier-Stokes equation:

$$\frac{\partial}{\partial t}(\rho \mathbf{v}) = -\boldsymbol{\nabla} \cdot \rho \mathbf{v}\mathbf{v} - \boldsymbol{\nabla} P - \boldsymbol{\nabla} \cdot \boldsymbol{\tau} + \rho \mathbf{g} \qquad (14)$$

where $\mu(\mathbf{x}) = \mu_0 + \mu_{sf}(\mathbf{x})$.

Eq. (14) represents a simply but efficient modification over the conventional Navier-Stokes equation. It incorporates a new viscosity of the fluid-solid interactions. According to Eq. (14), the fluid system has a viscosity that spatially varies due to the fluid's interactions with the solid surface. This modified Navier-Stokes equation allows the implementation of the no-slip boundary conditions, and the flawed slip boundary conditions are skipped.

Based on the hypothesis that we proposed in Section 2, the viscosity $\mu_{sf}(\mathbf{x})$ varies from one fluid layer to the other depending on the distance between the fluid layer and the solid surface. This gives $\mu_{sf}(\mathbf{x})$ a continuous-spatially varying function. Despite this fact, factors such as the wettability of the solid surface would make the fluid viscosity at the interfacial region is totally different than its viscosity elsewhere. For instance, the fluid particles would be depleted from the interfacial region if the fluid approaches a hydrophobic surface[16,18,31–34,50,51]. In addition, observations of the multiphase structure of water in nanotubes have revealed the phase transition into vapor at the interface, ice at the first-few water layers, and liquid in the core[17,31,52–59]. A viscosity function that can be used to describe the multiphase structure of the fluid and the possibility of forming a depletion region at the interface is, therefore, a piecewise continuous function that distinguishes the interfacial viscosity from the viscosity of the fluid elsewhere.

Studies have been conducted on the viscosity of fluids that exhibit slip/stick behaviors over solid surfaces (these studies are reviewed in Table 2 (also see Ref.[60])). It was demonstrated that the fluid viscosity strongly depends on the confinement size (if the fluid is confined), the applied shear rate, the fluid's temperature, and the wettability of the solid surface. Different approaches have been adopted to determine the fluid viscosity dependent on these factors. In one approach, the fluid viscosity is assumed spatially constant, and the effective viscosity is determined by intersecting the conventional model of the fluid flow – which assumes no-slip boundary – with the obtained results by MD simulations or experimental measurements (see Group 1 in Table 2). In another approach, the fluid interface with the solid surface is considered of a distinct-arbitrary assumed viscosity, and the effective viscosity is determined by the weighted-average between the interfacial viscosity and the bulk viscosity ($\mu_0$) (see Group 2 in Table 2). These two approaches, however, cannot determine the spatial variation of the fluid viscosity. Other approaches, therefore, have been proposed to give the spatial variation of the viscosity. Nonetheless, these methods were limited to fluids over flat surfaces (see Group 3 in Table 2). For example, an experimental approach that depends on measuring the fluid's shear forces generated between the tip of the atomic force microscope (AFM) and the fixed surface at different separation distances was proposed[61,62]. The obtained results, when collected, would give a distribution of the viscosity as a function of the distance from the flat surface. Another approach of determining the spatial variations of water viscosity between two flat surfaces depended on MD simulations[63,64].

Table 2: Review of approaches of determining the viscosity of nanoconfined fluids.

| No. | Assumptions | Method | Ref. |
|---|---|---|---|
| **Group 1** | | | |
| 1 | • Water is confined between two mica crystals.<br>• Constant viscosity distribution between the two surfaces.<br>• Effective shear viscosity was determined depending on the frequency of the oscillation and the twist angle between the two surfaces. | EXP | 65 |
| 2 | • Methane flows in a CNT.<br>• Constant viscosity distribution inside the CNT.<br>• The viscosity was determined by fitting the velocity profiles. | MD | 66 |
| 3 | • Water flow in nanotube.<br>• Water viscosity is constant in the radial direction within the water core.<br>• A modified Hagen–Poiseuille model with slip boundary conditions was used to determine the viscosity. | EXP | 40 |
| 4 | • A monolayer of liquid water confined between two surfaces.<br>• An ice layer was formed.<br>• Reported density distributions between the confining surfaces.<br>• No viscosity was reported. | MD | 54 |
| 5 | • The liquid–vapor phase transition near a weakly attractive surface (hydrophobic pores).<br>• Reported density distributions.<br>• No viscosity was reported. | MD | 31 |
| 6 | • Nonequilibrium MD simulations of the flow of liquid–vapor water mixtures and mixtures of water and nitrogen inside carbon nanotubes.<br>• Reported density distributions.<br>• No viscosity was reported. | MD | 52 |
| 7 | • Water nanoconfined between two mica surfaces. | MD | 67 |

|    | | Method | Ref. |
|----|---|---|---|
|    | • Viscosity is constant between the mica surfaces.<br>• The viscosity was reported as a function of the shear rate. | | |
| 8  | • Water flow in CNTs.<br>• Viscosity has a constant distribution inside the CNT.<br>• Viscosity was reported as a function of the CNT diameter. | MD | 68 |
| 9  | • Water nanoconfined between two mica surfaces.<br>• Viscosity is constant between the two mica surfaces.<br>• The viscosity was reported as a function of the shear rate. | EXP | 19 |
| 10 | • Water confined between two hydrophilic surfaces.<br>• By assuming a constant viscosity distribution and based on Feibelman model of a sphere on plate geometry, the effective viscosity was related to the peak of the friction force at the interface. | EXP | 69 |
| 11 | • Reported the viscoelasticity of a thin film of aqueous NaCl solution confined between mica surfaces.<br>• The effective viscosity was determined depending on the damping constant of the water flow and depending on the effective area of the fringes of equal chromatic order. | EXP | 70 |
| 12 | • The viscosity of nanometer-thick water films at the interface with an amorphous silica surface was measured.<br>• The viscosity values were obtained from three different measurements.<br>• No viscosity distributions were reported. | EXP | 71 |
| 13 | • Water flow in CNTs.<br>• Reported viscosity as a function of the nanotube size and a function of the flow rate. | MD | 72 |
| 14 | • Water flow in CNTs.<br>• Assumed constant viscosity and a plug-like flow.<br>• Fitted the velocity profile. | MD | 73 |
| 15 | • Bi- and tri-layer water films confined between hydrophilic surfaces.<br>• Density distributions between the surfaces were reported.<br>• No viscosity was reported. | MD | 74,75 |
| 16 | • Water flow in CNTs.<br>• Reported the variation of the effective viscosity and slip length as a function of the CNT diameter.<br>• The viscosity is constant within water core.<br>• The effective viscosity was assumed a weighted-average of the viscosities in the interface and core regions in the CNT cross section. | MD | 45,49,76 |
| 17 | • Phase transitions in confined water nanofilms based on density radial distributions.<br>• No viscosity was reported. | MD | 59 |
| 18 | • Reported radial density distributions.<br>• Did not report viscosity radial distribution.<br>• The viscosity was considered constant within the nanopore.<br>• The viscosity was reported as a function of the temperature. | MD | 77 |
| **Group 2** | | | |
| 19 | • Water flow in CNTs.<br>• Viscosity was calculated based on the Eyring theory.<br>• Variation of effective viscosity with temperature.<br>• Variation of effective viscosity with nanotube diameter. | MD & Eyring theory. | 42 |
| 20 | • Water flow in CNTs.<br>• Effective viscosity was reported as function of the temperature and the nanotube size. | Eyring-MD | 78,79 |
| 21 | • The viscosity profiles were determined with an interfacial viscosity and a constant core viscosity. | MD | 80 |
| 22 | • The effective viscosity of the confined water is a weighted average of the viscosities in the interface and bulk-like regions in nanopores.<br>• Also, the viscosity was related to the surface contact angle. | | 14,15 |
| 23 | • The viscosity at the interface is different than the viscosity at the water core region. | MD | 18 |
| 24 | • Viscosity within the core is constant.<br>• Assumed different interfacial viscosity and core viscosity.<br>• The core viscosity is constant. | Continuum Mechanics | 81 |
| 25 | • Assumed a spatially averaged value of the viscosity in the nanochannel. | MD | 82 |
| **Group 3** | | | |
| 26 | • Water on a flat surface.<br>• Density distributions were reported.<br>• Effective viscosity was reported assuming a constant viscosity distribution between the tip and the surface.<br>• The viscosity was reported for different values of the separation between the tip and the distance. | EXP | 61 |
| 27 | • The viscosity was reported as a function of the separation size between the tip of the AFM and the surface.<br>• Constant viscosity was assumed.<br>• Also, the viscosity was determined as a function of the surface wettability. | EXP | 62 |
| 28 | • Did not report a viscosity distribution.<br>• The viscosity was reported as a function of the water density, which would give a radial distribution of the water viscosity if intersected with the radial density distribution. | MD | 83 |
| 29 | • Reported radial density distribution.<br>• Related the viscosity to the size of the CNT.<br>• The viscosity was reported as a function of the water density, which would give a radial distribution of the water viscosity if intersected with the radial density distribution. | MD | 84,85 |
| 30 | • The variation of the shear viscosity of confined water between two graphene layers. | MD | 63 |
| 31 | • Presented an algorithm for calculating the spatial variation of the shear viscosity and thermal conductivity through an equilibrium solid-liquid interface using the zero-flux version of the boundary fluctuation theory. | Theory | 64 |

Recently, the author developed an effective approach to determine the spatial variation of the viscosity for fluids confined in hydrophobic and hydrophilic nanotubes[16,17]. Here, we represent this approach in a generalized form to determine the viscosity function $\mu(\mathbf{x}) = \mu_0 + \mu_{sf}(\mathbf{x})$ for different fluid systems. Therefore, we propose the viscosity in the following form:

$$\mu(|\mathbf{x} - \mathbf{x}_s|) = \begin{cases} \mu_I = \beta/VPR & 0 \leq |\mathbf{x} - \mathbf{x}_s| \leq \delta \\ \mu_c = \mu_0(1 + \xi(|\mathbf{x} - \mathbf{x}_s|)) & |\mathbf{x} - \mathbf{x}_s| > \delta \end{cases} \quad (15)$$

where $|\mathbf{x} - \mathbf{x}_s|$ is the distance between the fluid layer and the solid surface, $\delta$ is the thickness of the fluid-solid interface region, VPR is a parameter that is the ratio of the slip velocity at the fluid-solid interface to the applied pressure/shear stress gradient that act on the fluid system, $\beta$ is a geometrical parameter, and $\xi(|\mathbf{x} - \mathbf{x}_s|) = \mu_{sf}(|\mathbf{x} - \mathbf{x}_s|)/\mu_0$.

In Eq. (15), the fluid viscosity at the interface (i.e., interfacial viscosity $\mu_I$) depends on the geometry of the solid surface that confine the fluid and the fluid's slip velocity over the surface. The VPR parameter, which quantifies the slip velocity to the applied pressure/shear stress gradient, can be determined by measuring the fluid's slip velocity over the solid surface for different applied pressure/shear stress gradients. The viscosity within the fluid core $\mu_c$ (apart from the interface) spatially varies depending on the nondimensional function $\xi(|\mathbf{x} - \mathbf{x}_s|)$. The function $\xi(|\mathbf{x} - \mathbf{x}_s|)$ can be determined as the ratio of the molecular force of the fluid-solid interaction to the molecular force of the fluid-fluid interaction[16,17].

**4.1. Application to fluid flow in nanotubes**

To show the applicability of the proposed approach of the HCMM, it is implemented to model the fluid flow in nanotubes. Consider the steady state-laminar flow of a fluid in a circular tube of length $L$ and radius $R$. The fluid flows under the influence of a constant pressure gradient, $p$. Under these conditions, the non-zero components of the velocity and the viscous stress are $v_z(r)$ and $\tau_{rz}(r)$, respectively (assuming the cylindrical coordinates $r, \theta, z$). Considering these two non-zero components, the Navier-Stokes equation (Eq. (14)) of this system becomes:

$$p - \frac{1}{r}\frac{d}{dr}(r\tau_{rz}) = 0 \tag{16}$$

where according to Eq. (13), the viscous stress is expressed in the form:

$$\tau_{rz} = -\mu(r)\frac{dv_z(r)}{dr} \tag{17}$$

where $\mu(r)$ is the fluid's viscosity, which radially varies because of the fluid-solid interactions.

The implementation of a radially distributing viscosity that accounts for the fluid-solid interactions allows the application of the conventional no-slip boundary conditions of the considered fluid-tube system:

$$\tau_{rz}(0) = \text{finite} \tag{18}$$

$$v_z(R) = 0 \tag{19}$$

A general solution of Eq. (16) can be written in the form:

$$\tau_{rz} = p\frac{r}{2} + \frac{C_1}{r} \tag{20}$$

According to the boundary condition (18), $C_1 = 0$ arbitrary. Eq. (17) is then substituted into Eq. (20), and the result is solved for the velocity. Accordingly, the velocity is obtained after applying the boundary condition (19), as follows:

$$v_z(r) = \frac{p}{2}\left(\left[\int \frac{r}{\mu(r)}dr\right]_{r=R} - \int \frac{r}{\mu(r)}dr\right) \tag{21}$$

If the fluid-solid interactions are eliminated, Eq. (21) reduces to the velocity function of the conventional Hagen-Poiseuille model:

$$v_z^{HP}(r) = \frac{p}{4\mu_0}(R^2 - r^2) \tag{22}$$

This indicates an enhancement in the fluid volume flow rate by a factor $\epsilon$:

$$\epsilon = \frac{\int_0^R rv_z dr}{\int_0^R rv_z^{HP} dr} = \frac{2\mu_0}{R^4}\left(R\left[\int \frac{r}{\mu(r)}dr\right]_{r=R} - \int_0^R \left[r\int \frac{r}{\mu(r)}dr\right]dr\right) \tag{23}$$

*4.1.1. Viscosity calculations: water flow in nanotubes*

The radial variation of the viscosity is determined according to Eq. (15) for water flow in nanotubes. The fluid-solid interactions are defined by Lennard-Jones potential (as usually implemented in MD simulations[18,22,72,73]):

$$\varphi(r) = 4\epsilon_{sf}\left(\left(\frac{\sigma_{sf}}{R-r}\right)^{12} - \left(\frac{\sigma_{sf}}{R-r}\right)^{6}\right) \qquad (24)$$

where $\sigma_{sf}$ and $\epsilon_{sf}$ are the constants of Lennard-Jones potential of the fluid-solid interactions. The parameters $\beta$ and VPR of the interfacial viscosity $\mu_I$ are determined. First, the thickness of the interfacial region is defined by $\delta = 1.1224\,\sigma_{sf}$, which is the distance from the solid surface at which the fluid–solid interaction force $(d\varphi(r)/dr)$ is zero[16]. Because $\delta$ is too small, to a great extent, it is acceptable to assume a constant viscosity of the interfacial region. Then, the slip velocity at the first fluid layer is calculated according to Eq. (21), i.e., $v_s = v_z(R - \delta)$, in terms of the interfacial viscosity $\mu_I$, as follows:

$$v_s = \frac{p(R^2 - (R - \delta)^2)}{4\mu_I} \qquad (25)$$

Comparing Eq. (25) to Eq. (15), the geometrical parameter $\beta$ is determined by:

$$\beta = \frac{R^2 - (R - \delta)^2}{4} \qquad (26)$$

and the parameter VPR is, then, the ratio of the slip velocity to the applied pressure gradient, i.e., VPR = $v_s/p$.

The VPR parameter can be determined by a set of experimental measurements or MD simulations that can determine the slip velocity as a function of the applied pressure gradient. Thus, the VPR parameter is the slope of the $v_s - p$ relation. Based on observations from the literature and the results of experiments and MD simulations [6–8,22,44,45,73,86], the author determined the VPR parameter for water flow in hydrophilic and hydrophobic nanotubes[16,17]. Fig. 4 shows the VPR parameter as a function of the nanotube radius $R = 0.65 \to 15$ nm and for different values of the fluid-solid interaction energy $\epsilon_{sf} = 0.1 \to 3$ kJ/mol.

As for the core viscosity, it radially varies according to Eq. (15). The function $\xi(r)$ is determined as the ratio of the molecular force of the fluid-solid interaction $(d\varphi(r)/dr)$ to the maximum molecular force of the fluid-fluid interaction $(2.4\epsilon_{ff}/\sigma_{ff})$[16,17]:

$$\xi(r) = -\frac{5\epsilon_{sf}\sigma_{ff}}{3\epsilon_{ff}\sigma_{sf}}\left(12\left(\frac{\sigma_{sf}}{R-r}\right)^{13} - 6\left(\frac{\sigma_{sf}}{R-r}\right)^{7}\right) \qquad (27)$$

where $\sigma_{ff}$ and $\epsilon_{ff}$ are the constants of Lennard-Jones potential of the fluid-fluid interactions.

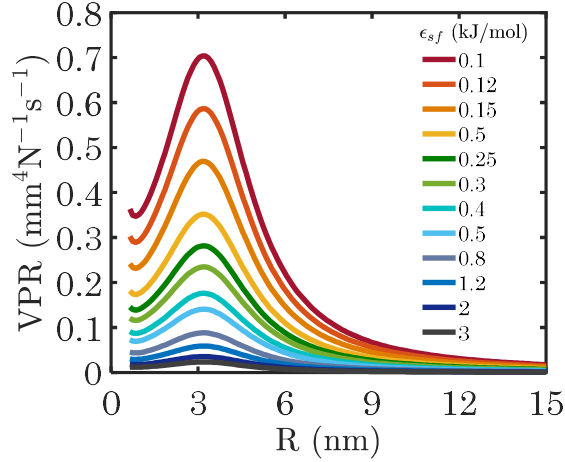

**Figure 4:** The VPR parameter of water flow in nanotubes. It is determined as a function of the nanotube radius $R$ for different values of the fluid-solid interaction energy $\epsilon_{sf}$.

## 5. Model Verification

To show the effectiveness of the proposed HCMM to reveal the fluid slip flow, it is tested against experimental models and MD simulations of over 90 different cases from the literature. We present in Table 3 the results of the proposed HCMM in comparison to the results of the experiments on water flow in different nanoporous membranes and different nanotubes[1,2,6–8]. In addition, we present in Fig. 5 the results of the proposed HCMM for water flow in different nanotubes in comparison to the results of the experiments and MD simulations as reported in Refs.[6,22,45,73,87]. The proposed model revealed an excellent match with the experiments and MD simulations.

**Table 3:** Flow enhancement factors of water flow in various nanoporous membranes. Comparison between the proposed model and experiments available in the literature.

| Membrane/Nanotube | Nanopore/Nanotube Diameter (nm) | Enhancement Factor ($\epsilon$) Experiment | Enhancement Factor ($\epsilon$) HCMM | Ref. |
|---|---|---|---|---|
| CNT – Membrane | 42 | 22 | 4 | 2 |
| MWCNT – Membrane | 7 | 1120-4073 | 162 - 2206 | 8 |
| DWCNT – Membrane | 1.3 – 2 | 713 - 4000 | 290 - 4280 | 7 |
| Polycarbonate Membrane | 15 | 3.7 | 1.61-1.77 | 7 |
| CNT | 0.81 | 882 | 793 | 6 |
| CNT | 0.87 | 662 | 487 | 6 |
| CNT | 0.98 | 354 | 255 | 6 |
| CNT | 1.42 | 103 | 75 | 6 |
| CNT | 1.52 | 59 | 60 | 6 |
| CNT | 1.59 | 51 | 54 | 6 |
| Boron Nitride NT | 46 | No slip | 0.98 | 1 |
| Boron Nitride NT | 52 | No slip | 0.95 | 1 |

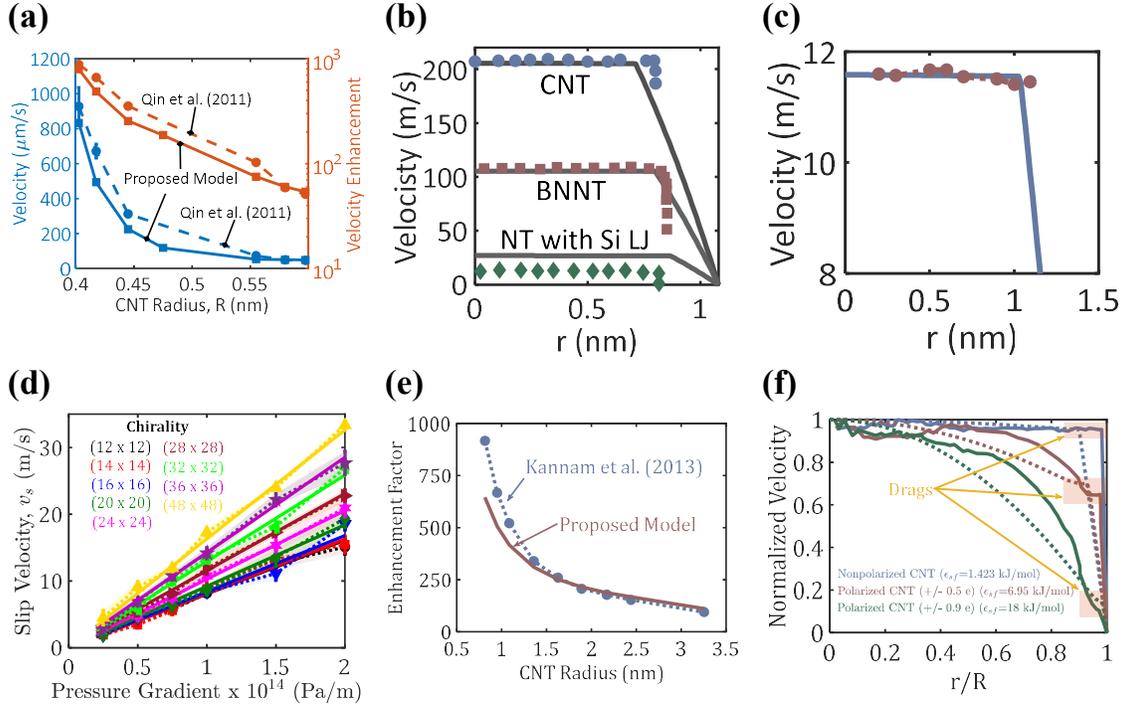

**Figure 5:** Comparison of the results of the proposed model with the results of experiments and MD simulations for water flow in nanotubes. (a) The velocity and the velocity enhancement factor of water flow in CNTs as functions of the nanotube radius in comparison to Qin et al.[6]. Using an array of field effect transistors, Qin et al.[6] measured the enhancement in the velocity (ratio of the measured velocity to the Hagen–Poiseuille velocity) of water flow on ultra-long CNT of 1330 μm length. Due to an electrical field that is used to flow the water on the CNT, the flow of water is damped over a damping length of 280 μm. To account for the damping effect, an equivalent length of the CNT of 0.025 m is considered in the performed analyses. (b) Velocity profiles of water flow in CNT, boron nitride nanotube (BNNT), and silicon nanotube (NT with LJ are of Si) in comparison to Joseph and Aluru et al.[73]. The applied pressure gradient is $p = 2 \pm 0.4 \times 10^{15}$ Pa/m and the nanotube radius is $R = 1.08$ nm. The results of the proposed model are represented by solid curves. (c) Velocity profile of water flow in a CNT with radius of $R = 1.385$ nm for an applied pressure gradient of $p = 1.24 \times 10^{14}$ Pa/m in comparison to Thomas and McGaughey[45]. The results of the proposed model are represented by the solid curve. (d) Slip velocity as a function of the applied pressure gradient for water flow in CNTs. Symbols with error bars and shaded regions refer to the results of the MD simulations by Kannam et al.[22] while solid lines are the results of the proposed model. (e) Flow enhancement factor as a function of the CNT radius as obtained by the proposed model (solid curve) in comparison to the results of MD simulations obtained by Kannam et al.[22] (doted curve). (f) Results of the proposed model for water flow in nonpolarized and polarized CNTs in comparison to the results of MD simulations obtained by Majumder and Corry[87]. Velocity profiles are normalized with respect to the corresponding maximum velocity.

The majority of the experiments of fluid flow in nanotubes depended on measuring the flow characteristics by collecting the fluid's permeate[2,7,8]. For instance, by measuring the pressure and the permeate of water flow for different imposed flow rates through 76 μm-thick nanoporous membrane with $1.07 \times 10^{10}$ #/cm² CNTs of $43 \pm 3$ nm diameters, Whitby et al.[2] determined the flow rate per the single CNT by $\sim(0.39 \to 2) \times 10^{-19}$ m³/s. In another experiment, Majumder et al.[8] measured the permeability of 34 → 126 µm – thick nanoporous membranes with $5 \times 10^{10}$ #/cm$^2$ multi-walled carbon nanotubes of 7 nm diameter, and the flow rate per a single pore was calculated by $\sim(2 \to 3.4) \times 10^{-19}$ m$^3$/s. For water flow through 2 → 3 µm – thick nanoporous membranes with $\leq 0.25 \times 10^{12}$ #/cm$^2$ double-walled CNTs (DWCNTs) of 1.3 → 2 nm diameters, Holt et al.[7] measured the flow rate per the single CNT by $\sim(0.77 \to 3) \times 10^{-20}$ m$^3$/s. The reported flow rates by these experiments were higher than the conventional no-slip flow rates of water in tubes of the same sizes. The factors of the flow enhancement as reported by these experiments and others are summarized in Table 3. We reconsidered the same experiments using the proposed HCMM, and the results of the HCMM are compared to the experimental results in Table 3. The results of the HCMM are in a good agreement with these experiments.

Based on the proposed hypothesis and the HCMM, the water flow in hydrophobic nanotubes is characterized by a velocity jump at the interface (Figs. 5(b), 5(c), and 5(f)). This is mainly because the water particles are weakly held to the solid surface, and the applied impulse is higher than the drag due to the water-solid interactions. Thus, the magnitude of the slip velocity at the interface depends on the applied pressure gradient, as it increases due to an increase in the applied pressure gradient (Fig. 5(d)). In addition, the slip velocity robustly depends on the strength of the water-solid interactions where it decreases as the strength increases (Fig. 5(b)).

Despite water would experience a slip at the interface with a hydrophobic surface, water flow would be significantly inhibited at the first water layers due to high polarized water-solid interactions (Fig. 5(f)). For instance, water flow in nonpolarized CNTs was obtained with a velocity jump at the interface followed by a plug-like flow, as shown in Fig. 5(f). For such a case, water is weakly held to the CNT surface because of a low strength of the water-solid interactions ($\epsilon = 1.423$ kJ/mol). When polarized CNTs were used, water became tightly linked to the CNT, and its flow is dragged at the first water layers after the interface. The water drag is sustained for a few layers after the interface, and it is reduced at other layers. As the strength of the polarized interaction energy is increased, the slip velocity is decreased while the drag at the first water layers is increased, as shown in Fig. 5(f).

For water flow in CNTs, the factor by which the flow is enhanced increases as the nanotube radius decreases (Figs. 5(a) and 5(e)). This size effect is mainly because of the surface wettability of the nanotube. The effect of the hydrophobicity of the nanotube increases as its radius decreases, and water slip is much promoted as the nanotube radius decreases. These factors make water flow is enhanced when confined in nanotubes with small sizes. In addition, these observations indicate that water flow converges to the no-slip water flow as the nanotube size increases.

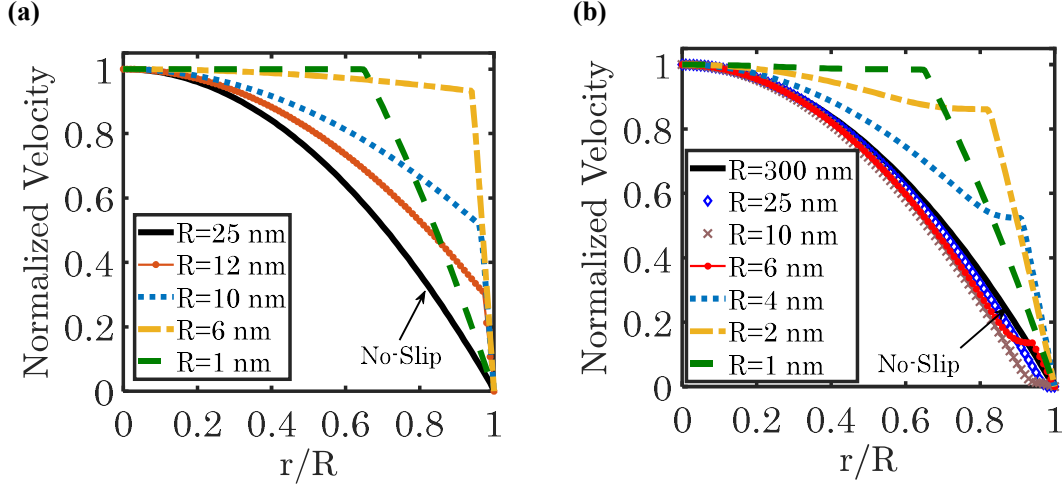

**Figure 6:** Normalized velocity profiles of water flow in nanotubes (results of the HCMM). The normalized velocity (normalized with respect to the corresponding maximum velocity) versus the nondimensional radial position ($r/R$) of water flow in nanotubes. The strength of the water-solid interaction is (a) $\epsilon_{sf} = 0.2$ kJ/mol and (b) $\epsilon_{sf} = 20$ kJ/mol.

We present in Fig. 6 the evolution of the conventional no-slip nondimensional velocity profile of water flow in circular tubes to anomalous nondimensional velocity profiles as the size of the tube decreases. The features of the velocity profile strongly depend on the strength of the water-solid interactions. Therefore, the results are extracted for $\epsilon_{sf} = 0.2$ kJ/mol and $\epsilon_{sf} = 20$ kJ/mol. When $\epsilon_{sf} = 0.2$ kJ/mol, water particles are weakly held to the tube, and the tube surface is hydrophobic. Under these weak water-solid interactions, water flow in a nanotube with a radius lower than 25 nm is featured with a velocity jump at the interface followed by a nearly-parabolic flow (Fig. 6(a)). The magnitude of the slip velocity decreases as the tube radius increases. The velocity profile is identical to the no-slip velocity profile when the nanotube radius is 25 nm or higher. When $\epsilon_{sf} = 20$ kJ/mol, water particles are tightly linked to the tube, and a drag would be observed at the first fluid layers (Fig. 6(b)). Under these strong water-solid interactions, water flow in a nanotube with a radius less than 10 nm is featured with a velocity jump at the interface, a severe drag at the first fluid layer, and a nearly-parabolic flow in the core. The slip velocity vanishes when the nanotube radius is $\geq 10$ nm. Nonetheless, the drag at the first water layers is sustained for water flow in a nanotube with a radius lower than 300 nm. For this case of the strong water-solid interactions, the conventional no-slip flow is obtained for water in a tube with a radius $\geq 300$ nm. These results indicate that the fluid flow and its slip characteristics strongly depend on its interactions with the solid surface. These interactions robustly impact the fluid flow in nanostructure such as nano-channels and nanotubes. In other cases, however, the fluid flow is conventional and the role of these interactions is negligible.

## Conclusions

We demonstrated that the excess molecular interactions between the fluid particles and the solid particles would significantly influence the fluid core as well as its boundary. This makes the correction of the classical fluid mechanics for the fluid slip by the slip boundary conditions is not an accurate approach. The slip boundary conditions cannot model the drag of the different fluid layers due to the fluid-solid interactions. Therefore, we proposed the correction of the classical fluid mechanics for the fluid slip by means of the hybrid continuum-molecular modeling. We put the bases of this new approach, which deals with the fluid as a continuous liquid matter that undergoes various forms of interactions with the solid surface. These interactions exceed the fluid boundary and significantly alter the fluid's viscosity. The implementation of the hybrid continuum-molecular modeling for water flow in nanotubes indicated an efficient approach that can model the slip flow of various fluid systems.

## Acknowledgements

The author acknowledges the research support from Abu Dhabi University (Grants 19300474 and 19300475).

## Competing Interests

The author has no competing interests to disclose.

## References

1. Secchi, E. *et al.* Massive radius-dependent flow slippage in carbon nanotubes. *Nature* **537**, 210–213 (2016).
2. Whitby, M., Cagnon, L., Thanou, M. & Quirke, N. Enhanced fluid flow through nanoscale carbon pipes. *Nano Lett.* **8**, 2632–2637 (2008).
3. Zhu, Y. & Granick, S. Limits of the Hydrodynamic No-Slip Boundary Condition. *Phys. Rev. Lett.* **88**, 106102 (2002).
4. Henry, C. L., Neto, C., Evans, D. R., Biggs, S. & Craig, V. S. J. The effect of surfactant adsorption on liquid boundary slippage. in *Physica A: Statistical Mechanics and its Applications* **339**, 60–65 (2004).
5. Schnell, E. Slippage of water over nonwettable surfaces. *J. Appl. Phys.* **27**, 1149–1152 (1956).
6. Qin, X., Yuan, Q., Zhao, Y., Xie, S. & Liu, Z. Measurement of the Rate of Water Translocation through Carbon Nanotubes. *Nano Lett.* **11**, 2173–2177 (2011).
7. Holt, J. K. *et al.* Fast mass transport through sub – 2-nanometer carbon nanotubes. *Science (80-. )*. **312**, 1034–1037 (2006).
8. Majumder, M., Chopra, N., Andrews, R. & Hinds, B. J. Nanoscale hydrodynamics: Enhanced flow in carbon nanotubes. *Nature* **438**, 44 (2005).
9. Joseph, P. & Tabeling, P. Direct measurement of the apparent slip length. *Phys. Rev. E - Stat. Nonlinear, Soft Matter Phys.* **71**, 035303(R) (2005).
10. Tretheway, D. C. & Meinhart, C. D. Apparent fluid slip at hydrophobic microchannel walls. *Phys. Fluids* **14**, L9–L12 (2002).
11. Lumma, D. *et al.* Flow profile near a wall measured by double-focus fluorescence cross-correlation. *Phys. Rev. E - Stat. Nonlinear, Soft Matter Phys.* **67**, 056313 (2003).
12. Cottin-Bizonne, C. *et al.* Nanorheology: An investigation of the boundary condition at hydrophobic and hydrophilic interfaces. *Eur. Phys. J. E* **9**, 47–53 (2002).
13. Zhu, Y. & Granick, S. Apparent slip of Newtonian fluids past adsorbed polymer layers. *Macromolecules* **35**, 4658–4663 (2002).
14. Wu, K. *et al.* Wettability effect on nanoconfined water flow. *PNAS* **114**, 3358–3363 (2017).
15. Wu, K. *et al.* Manipulating the Flow of Nanoconfined Water by Temperature Stimulation. *Angew. Chemie - Int. Ed.* **57**, 8432–8437 (2018).
16. Shaat, M. Viscosity of Water Interfaces with Hydrophobic Nanopores: Application to Water Flow in

17. Shaat, M. & Yongmei, Z. Fluidity and phase transitions of water in hydrophobic and hydrophilic nanotubes. *Sci. Rep.* **9**, 5689 (2019).
18. Sendner, C., Horinek, D., Bocquet, L. & Netz, R. R. Interfacial water at hydrophobic and hydrophilic surfaces: Slip, viscosity, and diffusion. *Langmuir* **25**, 10768–10781 (2009).
19. Leng, Y. & Cummings, P. T. Shear dynamics of hydration layers. *J. Chem. Phys.* **125**, 1–10 (2006).
20. Craig, V. S. J., Neto, C. & Williams, D. R. M. Shear-Dependent boundary slip in an aqueous newtonian liquid. *Phys. Rev. Lett.* **87**, 54504 (2001).
21. Zhu, Y. & Granick, S. Rate-Dependent Slip of Newtonian Liquid at Smooth Surfaces. *Phys. Rev. Lett.* **87**, 96105 (2001).
22. Kannam, S. K., Todd, B. D., Hansen, J. S. & Daivis, P. J. How fast does water flow in carbon nanotubes? *J. Chem. Phys.* **138**, 094701 (2013).
23. Kannam, S. K., Daivis, P. J. & Todd, B. D. Modeling slip and flow enhancement of water in carbon nanotubes. *MRS Bull.* **42**, 283–288 (2017).
24. Maxwell, J. C. On stresses in rarefied gases arising from inequalities of temperature. *Proc. R. Soc. London* **27**, 304–308 (1878).
25. Navier, C. L. Memorie sur les lois du lois du mouvement des fluides. *Mem. Acad. Sci Inst. Fr.* **6**, 298–440 (1827).
26. Lauga E., Brenner M., S. H. Microfluidics: The No-Slip Boundary Condition. in *Tropea C., Yarin A.L., Foss J.F. (eds) Springer Handbook of Experimental Fluid Mechanics. Springer Handbooks* (Springer, 2007).
27. Lamb, H. *Hydrodynamics*. (1932).
28. Batchelor, G. K. *An Introduction to Fluid Dynamics*. (Cambridge Univ. Press, 1967).
29. Israelachvili, J. N. Measurement of the viscosity of liquids in very thin films. *J. Colloid Interface Sci.* **110**, 263–271 (1986).
30. Chan, D. Y. C. & Horn, R. G. The drainage of thin liquid films between solid surfaces. *J. Chem. Phys.* **83**, 5311–5324 (1985).
31. Brovchenko, I., Geiger, A. & Oleinikova, A. Water in nanopores: II. The liquid-vapour phase transition near hydrophobic surfaces. *J. Phys. Condens. Matter* **16**, (2004).
32. Myers, T. G. Why are slip lengths so large in carbon nanotubes? *Microfluid. Nanofluidics* **10**, 1141–1145 (2011).
33. Chattopadhyay, S. *et al.* How water meets a very hydrophobic surface. *Phys. Rev. Lett.* **105**, 1–4 (2010).
34. Janeček, J. & Netz, R. R. Interfacial water at hydrophobic and hydrophilic surfaces: Depletion versus adsorption. *Langmuir* **23**, 8417–8429 (2007).
35. Zhang, T. *et al.* Mesoscopic method to study water flow in nanochannels with different wettability. *Phys. Rev. E* **102**, 13306 (2020).
36. Loose, W. & Hess, S. Rheology of dense model fluids via nonequilibrium molecular dynamics: Shear thinning and ordering transition. *Rheol. Acta* **28**, 91–101 (1989).
37. Cheng, J. T. & Giordano, N. Fluid flow through nanometer-scale channels. *Phys. Rev. E - Stat. Physics, Plasmas, Fluids, Relat. Interdiscip. Top.* **65**, 031206 (2002).
38. Maali, A. & Bhushan, B. Nanorheology and boundary slip in confined liquids using atomic force microscopy. *Journal of Physics Condensed Matter* **20**, 5052 (2008).
39. Churaev, N. V., Sobolev, V. D. & Somov, A. N. Slippage of liquids over lyophobic solid surfaces. *J. Colloid Interface Sci.* **97**, 574–581 (1984).
40. Choi, C. H., Westin, K. J. A. & Breuer, K. S. Apparent slip flows in hydrophilic and hydrophobic microchannels. *Phys. Fluids* **15**, 2897–2902 (2003).
41. Sinha, S., Pia Rossi, M., Mattia, D., Gogotsi, Y. & Bau, H. H. Induction and measurement of minute flow rates through nanopipes. *Phys. Fluids* **19**, 013603 (2007).
42. Babu, J. S. & Sathian, S. P. The role of activation energy and reduced viscosity on the enhancement of water flow through carbon nanotubes. *J. Chem. Phys.* **134**, (2011).
43. Kotsalis, E. Multiscale modeling and simulation of fullerenes in liquids. PhD Dissertation. (ETH, Zurich, 2006).
44. Falk, K., Sedlmeier, F., Joly, L., Netz, R. R. & Bocquet, L. Molecular origin of fast water transport in carbon nanotube membranes: Superlubricity versus curvature dependent friction. *Nano Lett.* **10**, 4067–4073 (2010).
45. Thomas, J. A. & Mcgaughey, A. J. H. Reassessing Fast Water Transport Through Carbon Nanotubes. *Nano Lett.* **8**, 2788–2793 (2008).
46. Gu, X. & Chen, M. Shape dependence of slip length on patterned hydrophobic surfaces. *Appl. Phys. Lett.* **99**, 063101 (2011).
47. Xiong, W. *et al.* Strain engineering water transport in graphene nanochannels. *Phys. Rev. E - Stat. Nonlinear, Soft Matter Phys.* **84**, 056329 (2011).
48. Ma, M. D. *et al.* Friction of water slipping in carbon nanotubes. *Phys. Rev. E - Stat. Nonlinear, Soft*

*Matter Phys.* **83**, 1–7 (2011).
49. Thomas, J. A., Mcgaughey, A. J. H. & Kuter-arnebeck, O. Pressure-driven water flow through carbon nanotubes: Insights from molecular dynamics simulation. *Int. J. Therm. Sci.* **49**, 281–289 (2010).
50. Doshi, D. A., Watkins, E. B., Israelachvili, J. N. & Majewski, J. Reduced water density at hydrophobic surfaces: Effect of dissolved gases. *Proc. Natl. Acad. Sci.* **102**, 9458–9462 (2005).
51. Sedlmeier, F. *et al.* Water at polar and nonpolar solid walls (Review). *Biointerphases* **3**, FC23–FC39 (2008).
52. Kotsalis, E. M., Walther, J. H. & Koumoutsakos, P. Multiphase water flow inside carbon nanotubes. *Int. J. Multiph. Flow* **30**, 995–1010 (2004).
53. Shiomi, J., Kimura, T. & Maruyama, S. Molecular dynamics of ice-nanotube formation inside carbon nanotubes. *J. Phys. Chem. C* **111**, 12188–12193 (2007).
54. Zangi, R. & Mark, A. E. Monolayer Ice. *Phys. Rev. Lett.* **91**, 1–4 (2003).
55. Giovambattista, N., Rossky, P. J. & Debenedetti, P. G. Phase Transitions Induced by Nanoconfinement in Liquid Water. *Phys. Rev. Lett.* **102**, 050603 (2009).
56. Kumar, H., Dasgupta, C. & Maiti, P. K. Phase Transition in Monolayer Water Confined in Janus Nanopore. *Langmuir* **34**, 12199–12205 (2018).
57. Raju, M., Duin, A. Van & Ihme, M. Phase transitions of ordered ice in graphene nanocapillaries and carbon nanotubes. *Sci. Rep.* **8**, 3851 (2018).
58. Cummings, P. T., Docherty, H. & Iacovella, C. R. Phase Transitions in Nanoconfined Fluids : The Evidence from Simulation and Theory. *AIChE J.* **56**, 842–848 (2010).
59. Han, S., Choi, M. Y., Kumar, P. & Stanley, H. E. Phase transitions in confined water nanofilms. *Nat. Phys.* **6**, 685–689 (2010).
60. Mahbubul, I. M., Saidur, R. & Amalina, M. A. Latest developments on the viscosity of nanofluid. *Int. J. Heat Mass Transf.* **55**, 874–885 (2012).
61. Li, T. De, Gao, J., Szoszkiewicz, R., Landman, U. & Riedo, E. Structured and viscous water in subnanometer gaps. *Phys. Rev. B - Condens. Matter Mater. Phys.* **75**, 1–6 (2007).
62. Ortiz-Young, D., Chiu, H. C., Kim, S., Voïtchovsky, K. & Riedo, E. The interplay between apparent viscosity and wettability in nanoconfined water. *Nat. Commun.* **4**, (2013).
63. Neek-amal, M., Peeters, F. M., Grigorieva, I. V & Geim, A. K. Commensurability Effects in Viscosity of Nanoconfined Water Commensurability Eeffects in Viscosity of Nanoconfined Water. (2016). doi:10.1021/acsnano.6b00187
64. Petravic, J. & Harrowell, P. Spatial dependence of viscosity and thermal conductivity through a planar interface. *J. Phys. Chem. B* **113**, 2059–2065 (2009).
65. Zhu, Y. & Granick, S. Viscosity of interfacial water. *Phys. Rev. Lett.* **87**, 1–4 (2001).
66. Sokhan, V. P., Nicholson, D. & Quirke, N. Fluid flow in nanopores: Accurate boundary conditions for carbon nanotubes. *J. Chem. Phys.* **117**, 8531–8539 (2002).
67. Leng, Y. & Cummings, P. T. Erratum: Fluidity of Hydration Layers Nanoconfined between Mica Surfaces (Physical Review Letters (2005) 94 (026101)). *Phys. Rev. Lett.* **94**, 19–22 (2005).
68. Liu, Y., Wang, Q., Wu, T. & Zhang, L. Fluid structure and transport properties of water inside carbon nanotubes. *J. Chem. Phys.* **123**, (2005).
69. Major, R. C., Houston, J. E., McGrath, M. J., Siepmann, J. I. & Zhu, X. Y. Viscous water meniscus under nanoconfinement. *Phys. Rev. Lett.* **96**, 5–8 (2006).
70. Sakuma, H., Otsuki, K. & Kurihara, K. Viscosity and lubricity of aqueous NaCl solution confined between mica surfaces studied by shear resonance measurement. *Phys. Rev. Lett.* **96**, 6–9 (2006).
71. Goertz, M. P., Houston, J. E. & Zhu, X. Y. Hydrophilicity and the viscosity of interfacial water. *Langmuir* **23**, 5491–5497 (2007).
72. Chen, X. *et al.* Nanoscale fluid transport: Size and rate effects. *Nano Lett.* **8**, 2988–2992 (2008).
73. Joseph, S. & Aluru, N. R. Why Are Carbon Nanotubes Fast Transporters of Water? *Nano Lett.* **8**, 452–458 (2008).
74. Pertsin, A. & Grunze, M. Quasistatic computer simulation study of the shear behavior of bi- and trilayer water films confined between model hydrophilic surfaces. *Langmuir* **24**, 4750–4755 (2008).
75. Pertsin, A. & Grunze, M. A Computer Simulation Study of Stick - Slip Transitions in Water Films Confined between Model Hydrophilic Surfaces . 1 . Monolayer Films. *Langmuir* 135–141 (2008).
76. Thomas, J. A. & McGaughey, A. J. H. Water flow in carbon nanotubes: Transition to subcontinuum transport. *Phys. Rev. Lett.* **102**, 1–4 (2009).
77. Xu, B. *et al.* Temperature dependence of fluid transport in nanopores. *J. Chem. Phys.* **136**, (2012).
78. Zhang, H., Ye, H., Zheng, Y. & Zhang, Z. Prediction of the viscosity of water confined in carbon nanotubes. *Microfluid. Nanofluidics* **10**, 403–414 (2011).
79. Ye, H., Zhang, H., Zhang, Z. & Zheng, Y. Size and temperature effects on the viscosity of water inside carbon nanotubes. *Nanoscale Res. Lett.* **6**, 87 (2011).
80. Bonthuis, D. J. & Netz, R. R. Beyond the continuum: How molecular solvent structure affects electrostatics and hydrodynamics at solid-electrolyte interfaces. *J. Phys. Chem. B* **117**, 11397–11413

(2013).
81. Mattia, D. & Calabro, F. Explaining high flow rate of water in carbon nanotubes via solid – liquid molecular interactions. 125–130 (2012). doi:10.1007/s10404-012-0949-z
82. Li, L., Kazoe, Y., Mawatari, K., Sugii, Y. & Kitamori, T. Viscosity and Wetting Property of Water Confined in Extended Nanospace Simultaneously Measured from Highly-Pressurized Meniscus Motion. *J. Phys. Chem. Lett.* **3**, 2447–2452 (2012).
83. Barati Farimani, A. & Aluru, N. R. Existence of Multiple Phases of Water at Nanotube Interfaces. *J. Phys. Chem. C* **120**, 23763–23771 (2016).
84. Köhler, M. H. & Da Silva, L. B. Size effects and the role of density on the viscosity of water confined in carbon nanotubes. *Chem. Phys. Lett.* **645**, 38–41 (2016).
85. Köhler, M. H., Bordin, J. R., Da Silva, L. B. & Barbosa, M. C. Breakdown of the Stokes-Einstein water transport through narrow hydrophobic nanotubes. *Phys. Chem. Chem. Phys.* **19**, 12921–12927 (2017).
86. Du, F., Qu, L., Xia, Z., Feng, L. & Dai, L. Membranes of vertically aligned superlong carbon nanotubes. *Langmuir* **27**, 8437–8443 (2011).
87. Majumder, M. & Corry, B. Anomalous decline of water transport in covalently modified carbon nanotube membranes. *Chem. Commun.* **47**, 7683–7685 (2011).